\documentstyle[11pt,aussois2003,twoside,psfig,epsf,graphicx]{article}
\def\pg{PG~1115+080}
\def\kmsmpc{km.s$^{-1}$.Mpc$^{-1}$}

\def\edcomment#1{\iffalse\marginpar{\raggedright\sl#1\/}\else\relax\fi}
\reversemarginpar
\begin{document}
%
\title{Quasar Lensing: the Observer's Point of View}
%
\author{F. Courbin}
\affil{Institut d'Astrophysique et de G\'eophysique, Universit\'e de 
Li\`ege,\\ All\'ee du 6 ao\^ut 17, Bat B5C, Li\`ege 1, Belgium}

\label{page:first}
\begin{abstract}
The determination of the Hubble parameter H$_0$ is probably one of the
most important  applications of quasar lensing.  The  method, based on
the  measurement of  the so-called  ``time-delay'' between  the lensed
images  of  distant  sources,  e.g.,  quasars, and  on  detailed  mass
modeling of  the potential well  responsible for the  multiple images,
yields an accuracy at least  comparable with other techniques and that
can be  improved further with  high precision observations, as  can be
obtained  with intrumentation of  constantly increasing  quality.  The
basics of the ``time-delay'' method are described, and the emphasis is
put on  the observational constraints available  to the astrophysicist
in order to  implement the method and to derive  an accurate value for
H$_0$, independent of  any standard candle or any  strong prior on the
other cosmological parameters.
\end{abstract}
\section{Why Observing Lensed Quasars ?}

Gravitational lensing is a  well established field of astrophysics. It
is well  enough understood that  it can be  applied to other  areas of
astrophysics  in order to  tackle astrophysical  problems under  a new
angle. Some  applications of gravitational lensing  concentrate on the
study of the objects responsible for the deflection of light, the {\it
lenses}.   Others,  aim  at  studying  the  stretched,  distorted  and
(de)magnified  images of  the background  objects, the  {\it sources}.
For example, stellar micro-lensing is used to probe the content of our
own  galaxy  in  dark  low-mass  stars,  or  micro-lenses.   The  weak
distortions of very distant galaxies  is used to detect indirectly and
even to  map what  might be  the largest lenses  in the  Universe: the
Large   Scale   Structures.    In   many  multiply   imaged   quasars,
(micro)lenses  are  found   within  (macro)lenses:  quasar  micro-  or
milli-lensing provides  us with information  on the structure  of both
the sources and the lenses (see for example Schechter 2003; Wambsganss
2003).

Many of  the applications of gravitational lensing,  and in particular
of  quasar  lensing,  were  known  and described  decades  ago.   They
nevertheless only start  now to be implemented on  a systematic basis,
taking  advantage of  the  recent  explosion of  the  number of  large
observatories that operate at high angular resolution and down to very
faint magnitudes.

Quasar  lensing helps  us to  study lenses  and sources,  but  it also
consists in a fantastic tool to study the space between the lenses and
the  sources  !   While the  light  travels  from  the source  to  the
observer, it is absorbed by the  Inter Stellar Medium of the lens, and
by  the  neutral gas  of  the  Inter-Galactic  Medium.  The  study  of
absorption  lines   in  multiply  imaged  quasars   provides  us  with
information about the geometry  of intergalactic clouds (Smette 2003).
Last, but not least, multiply imaged quasars tell us about the size of
the Universe, through the  measurement of the so-called ``time-delay''
between the lensed  images.  This quantity is directly  related to the
mass distribution  in the lensing  galaxy and to the  Hubble parameter
H$_0$.  The measurement of H$_0$  using lensed quasars is the topic of
the present chapter.

\section{First Discoveries and Searches}

\subsection{A Few Lucky Cases}

The observational  history of lensed  quasars starts with a  few lucky
cases found  ``by accident'' during surveys  or follow-up observations
of projects  unrelated to gravitational  lensing. The very  first case
was  the   double  quasar   Q~0957+561.   When  observed   at  optical
wavelengths,  the  z=1.405  quasar   appeared  as  two  point  sources
separated by 5.7\arcsec\ (Walsh  et al.  1979).  Spectra obtained with
the  Multi-Mirror-Telescope  showed   that  both  objects  had  almost
identical  spectral  properties   (Weymann  et  al.   1979),  strongly
supporting the hypothesis of  gravitational lensing: two images of one
single object were seen, due to the potential well created by a galaxy
along  the  line  of  sight.   In  fact, not  only  the  spectra  were
identical, but subtraction of the quasar images also revealed, for the
first time,  the lensing  galaxy, hidden by  the much  brighter quasar
images.  With such  observational  material, no  serious doubts  could
remain about the lensed nature of Q~0957+561.

Other cases were found soon after, such as the quadruply imaged quasar
PG~1115+080 (Weymann et al. 1980) that  we will use in this article to
illustrate how  lensed quasars can  help use to determine  H$_0$. Even
the famous  ``Einstein Cross'', Q~2237+0305 (Huchra et  al.  1985; see
Fig.  1),  was discovered during follow-up observations  in the course
of the CfA  redshift survey: a spectrum obtained  of the central parts
of  a  z=0.04  redshift  galaxy,  turned  out  to  display  the  exact
characteristics  of  a quasar  at  much  higher  redshift, z=1.7.   In
addition, the total apparent luminosity of the bulge of the galaxy was
far too high for a normal spiral.  Indeed, it was in fact the combined
light of the actual galaxy's bulge and of the four (unresolved) quasar
images.   High  resolution  images  taken  a few  years  later  nicely
confirmed that the object was  composed of four separate quasar images
almost  aligned with  the bulge  of  the spiral  galaxy (Schneider  et
al. 1988; see Fig. 2).

\begin{figure}[p!]
\begin{center}
\leavevmode
\includegraphics[height=5.8cm]{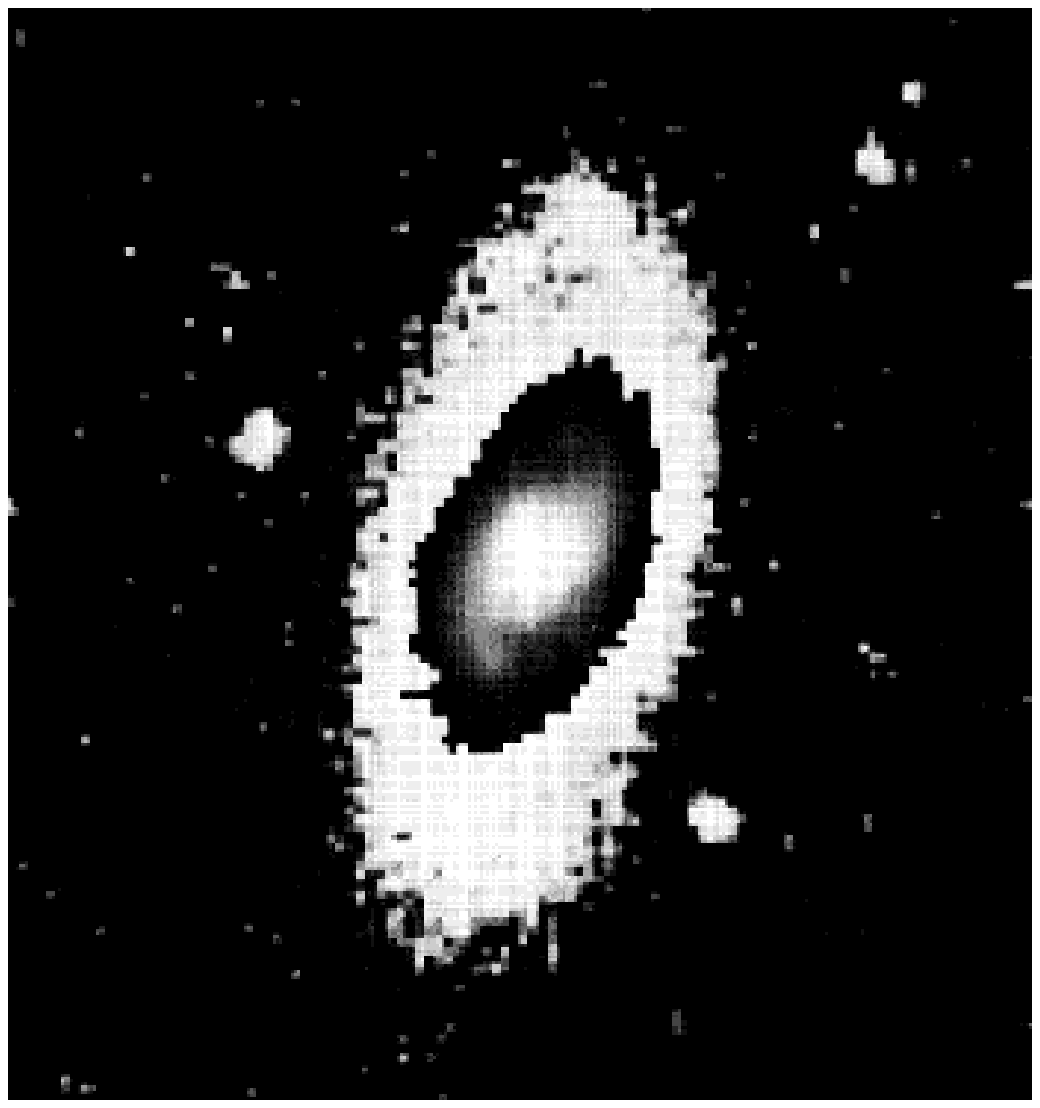}
\includegraphics[height=5.8cm]{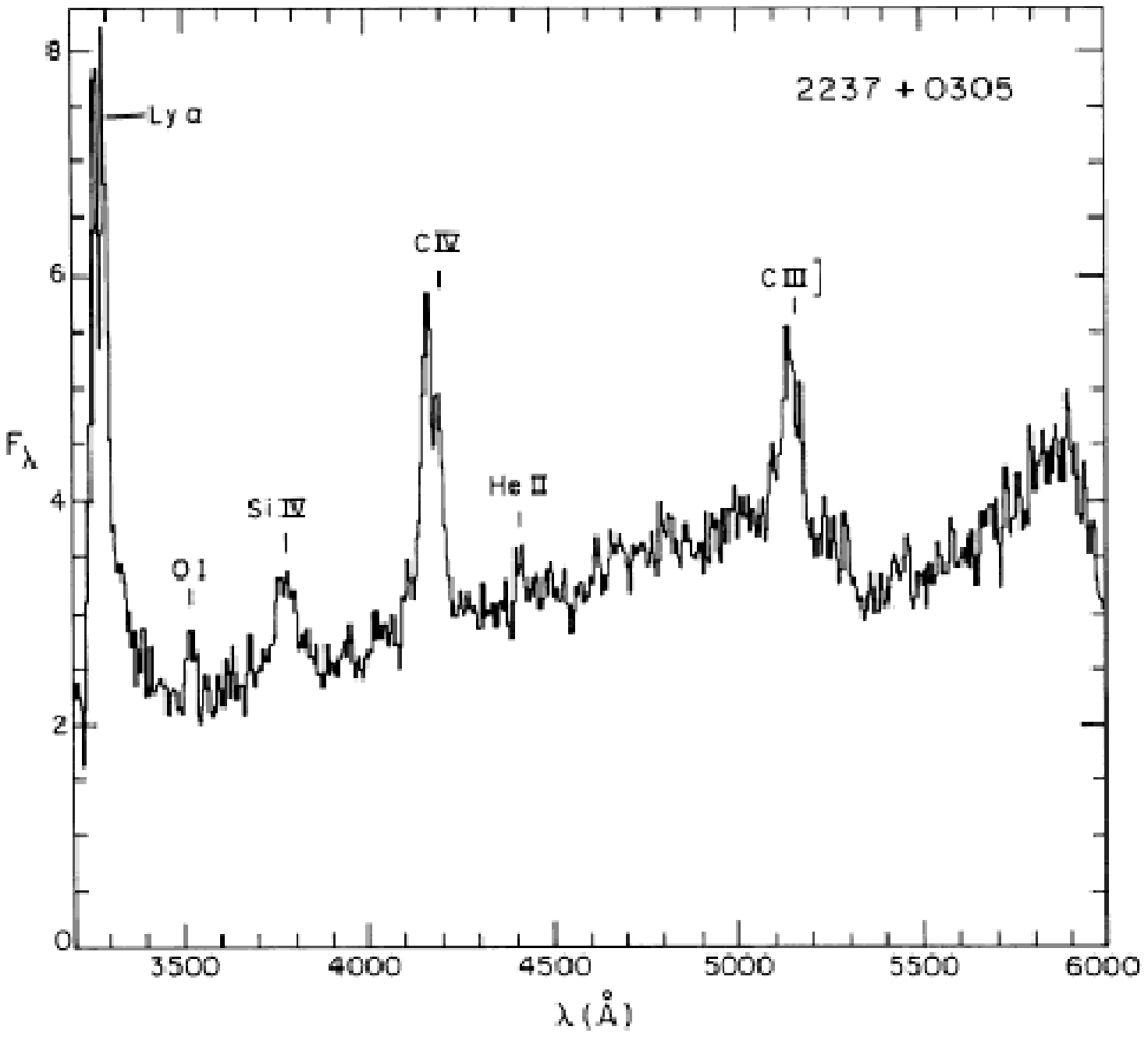}
\caption{{\it Left:} ground based image of the lensing galaxy (z=0.04)
in  the Einstein Cross.  The spatial  resolution is  low. It  does not
allow  to  discriminate  between  the  bulge of  the  galaxy  and  any
background quasar image(s). {\it  Right:} spectrum of the most central
part of  the galaxy.  The spectrum is
not the one of a low redshift galaxy, but that of a much more distant object:
a quasar at z=1.7 (Huchra et al. 1987).}
\label{EC_1}
\vspace*{5mm}
\leavevmode
\includegraphics[height=5.9cm]{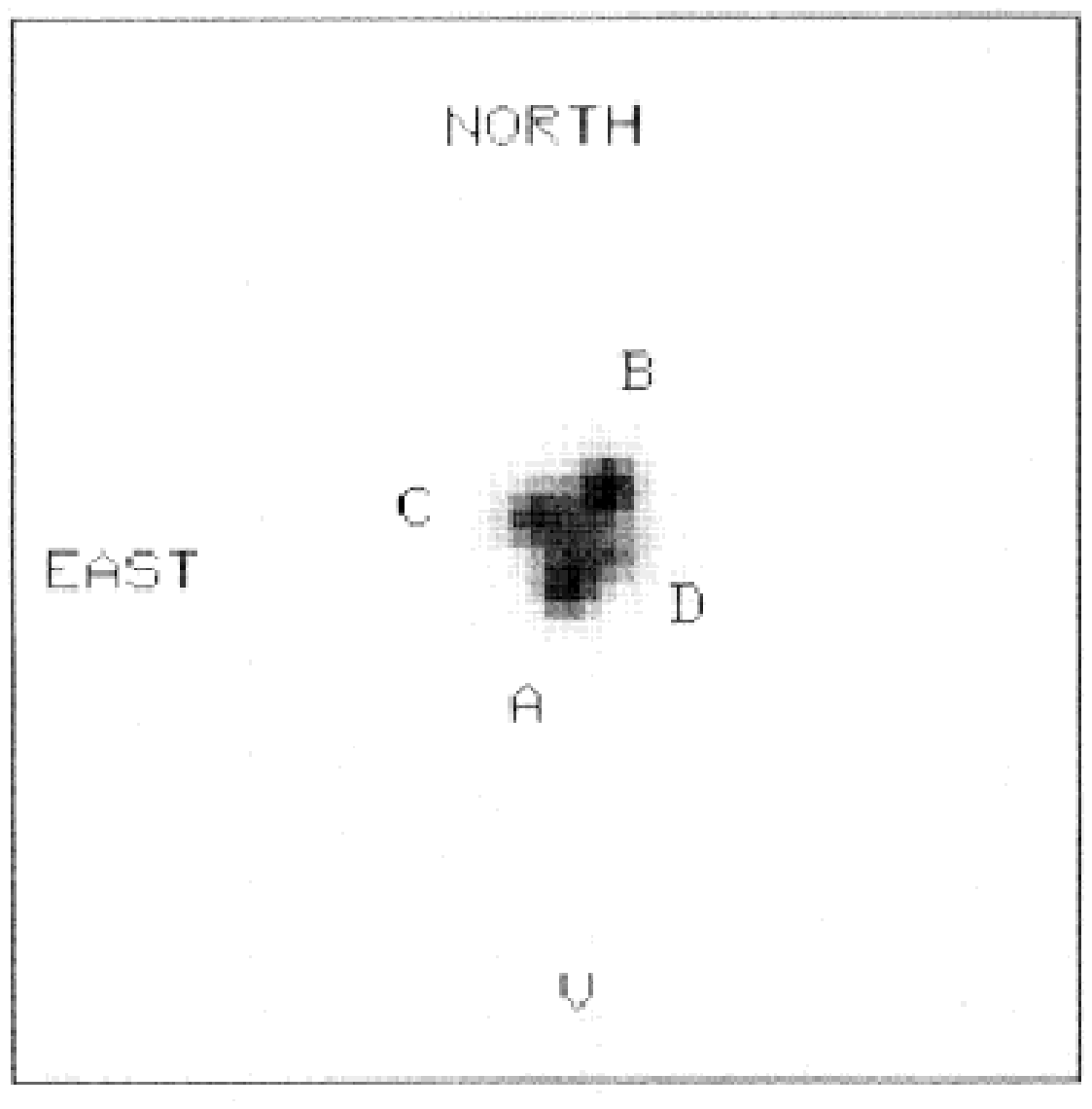}
\includegraphics[height=5.8cm]{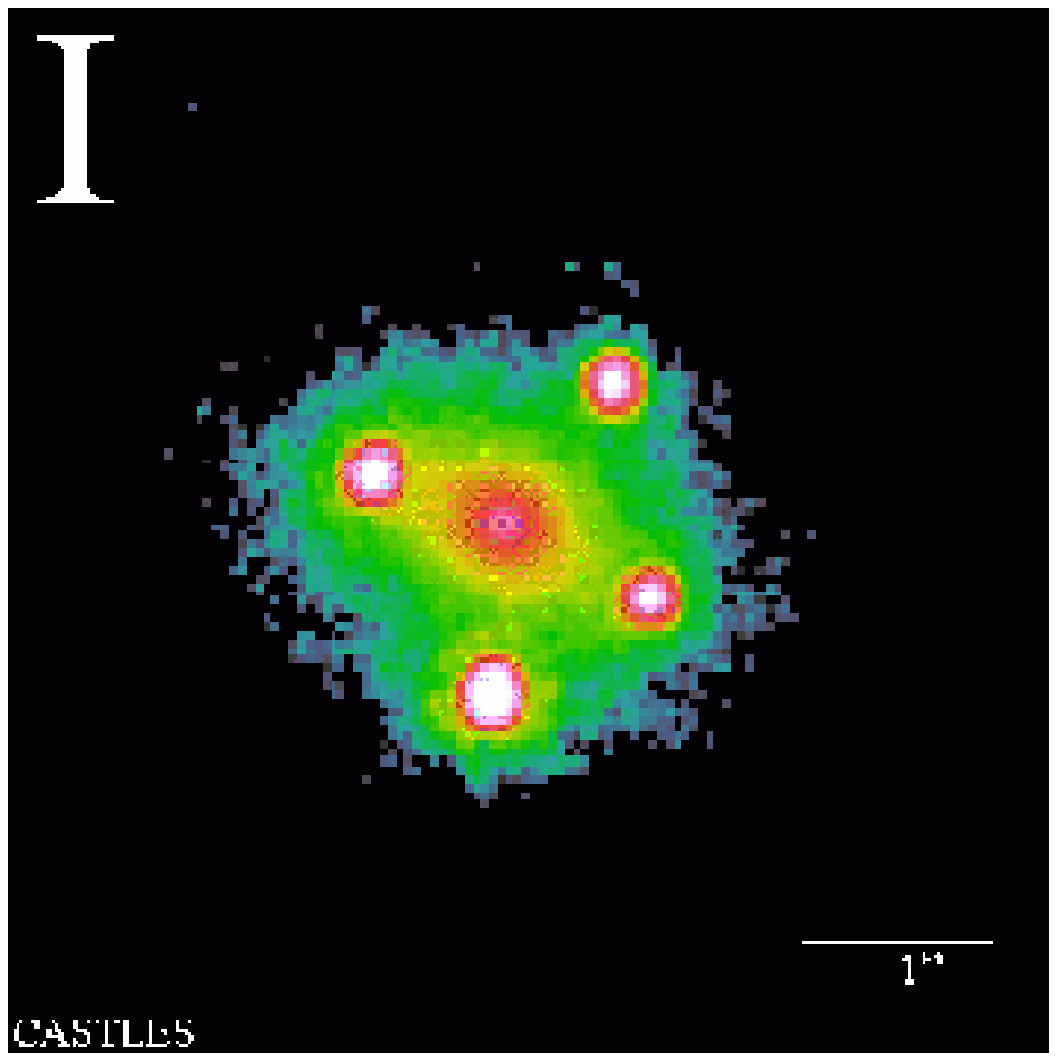}
\caption{{\it  Left:} ground  based  image of  the ``Einstein  cross''
unveiling, for the first time, four separated quasar images. The field
of view 10\arcsec wide (Schneider et al. 1988).  {\it Right:} HST view
of the Einstein cross, allowing for accurate astrometry and photometry
of the system  as well as detailed surface photometry  of the bulge of
the lensing galaxy.   The length of the white bar  at the bottom right
is 1\arcsec\ (Image taken from Kochanek et al. 2003a).}
\label{EC_2}
\end{center}
\end{figure}

\subsection{Systematic Searches and the Magnification Bias}

While the very first lenses were  found by chance (and there are still
lenses found by chance from time to time; see for example Sluse et al.
2003), observational strategies were  soon designed to find many more,
on purpose !

Such an enterprise  requires to estimate the number  of lensed objects
within a sample of quasars, given the selection criteria, in general a
flux limit.   The exercise  has been done  many times by  many groups,
following  the ideas first  proposed by  Turner (1980).   The original
idea of Turner was to describe  the effect of undetected lenses on the
apparent  evolution of  quasars.   As lensing  amplifies the  apparent
luminosity  of background  objects, significant  modifications  of the
observed luminosity function of  quasars were suspected, in particular
toward the bright end. Several ingredients are needed to carry out the
calculation: (1)  the spatial distribution of lenses,  (2) the spatial
distribution of sources, (3) a  realistic mass model for the lenses in
order to estimate the amplification, (4) a cosmological model and, (5)
the (unlensed)  luminosity function of the sources.   Although none of
these were precisely  known in the eighties, it  was quickly understood
that  lens statistics was  a particularly  sensitive function  of the
slope of the  source luminosity function: the relative  weights of the
faint  to bright  quasar  number  counts give  rise  to the  so-called
``magnification  bias'' (see  contribution  by Smette,  2003 for  more
details).   Its  net effect  on  a flux  limited  sample  is that  the
fraction of sources likely to be  magnified by a given amount $\mu$ is
higher for  apparently bright sources  than for fainter ones  (see for
example  Schneider, Ehlers  \& Falco,  1992). In  other  words, bright
sources are seen bright, because they are (more likely to be) lensed.

Based  on this  simple, but  important finding,  several  surveys were
started, targeting at quasars with the brightest absolute magnitude in
large  samples.  Because  at  the  time  of  the  first  surveys  were
initiated, high angular resolution was  easier to achieve in the radio
than in the  visible, multiply imaged quasars were  often the found in
large  radio  surveys  (e.g.,  Lawrence  et  al.   1986;  Langston  et
al. 1989;  Hewitt et al. 1992).   The largest of these  surveys so far
are probably CLASS,  the ``Cosmic Lens All Sky  Survey'' and JVAS, the
``Jorell Bank - VLA  Astrometric Surveys'' (see, among others, Patnaik
et al. 1992; Myers et al. 1995; Jackson et al. 1998).

Almost in  parallel with radio  surveys, optical searches  started and
successfully yielded a significant harvest of multiply imaged quasars.
Among the first ones to be  discovered, were the double UM 673 (Surdej
et al.  1987), the  quadruple ``cloverleaf'' H~1413+117 (Magain et al.
1988), followed  by some  cases in the  Hamburg/ESO survey  for bright
quasars (e.g.,  Wisotzki et al.  1993).   The search for  new cases is
still ongoing,  with a  success rate that  makes it difficult  to keep
track  of every  new  discovery.  Large  multi-wavelengths wide  field
surveys are  now relatively  easy to carry  out and  the ``multiband''
magnification bias  is studied  in order to  understand the  effect of
lensing on  such data  (Wyithe et al.   2003).  The prospects  to find
many  new  lenses suitable  for  cosmological  applications (e.g.,  in
SLOAN, FIRST, GOODS, etc...) are  therefore excellent.  At the time of
the writing  of this article,  several new good candidates  from these
surveys are under analyze.

\section{Lensed Quasars and $H_0$}

Quasar lensing helps to  solves astrophysical puzzles in various ways,
but  one   of  its  most   beautiful  applications  is   probably  the
determination of the Hubble parameter H$_0$.

\subsection{The Time-delay Method}

In 1964,  the Norwegian astronomer  Sjur Refsdal proposed  an original
me\-thod  (Refsdal 1964)  to use  gravitational lensing  as a  tool to
measure the size/age  of the Universe.  When photons  propagate from a
distant source toward  the observer, they are under  the effect of the
gravity field of lenses along the  line of sight. They do not follow a
straight line anymore, but their  trajectory is curved and longer than
the  original one.   As  a consequence,  it  takes more  time for  the
photons to travel from a lensed  source than from an unlensed one. The
{\it geometrical}  difference introduced by the lens  between the two,
lensed and  unlensed paths, introduces a time-lag  between the arrival
times  of the (lensed  and unlensed)  photons at  the position  of the
observer.   This time  lag is  called the  geometrical ``time-delay'',
$t_{\rm  geom}$.   While passing  in  the  immediate  vicinity of  the
gravity field  of the lens, the  light is affected by  a second delay:
the  gravitational time-delay,  $t_{\rm grav}$.   A  ``lensed photon''
will be  seen by an  observer with a  total time-delay $t_{\rm  tot} =
t_{\rm geom} +  t_{\rm grav}$, with respect to  the observation of the
same photon if it were not lensed.

The time-delay  is a function of  image position in  projection on the
plane of the  sky.  One can then define an  {\it arrival time surface}
that associates, to  each position on the sky, a given  a value of the
time-delay.  Most  of this surface is  missed by the  observer who has
only access to  the few areas where the lensed  images form.  When two
or more  images of the source  are observed it is  possible to compare
the  arrival  times at  the  positions of  the  lensed  images and  to
determine a ``relative  time-delay''.  This is in fact  the only truly
measurable  quantity, rather  than the  actual time-delay  between the
lensed and unlensed paths to  the source, since the unlensed source is
never visible.

In  practice, time-delays are  measured taking  advantage of  a lensed
source   with  significant   photometric  variations.    Due   to  the
time-delay,  the  variations  will  be  detected by  the  observer  at
different times in the light curves  of each image.  The shift in time
between the light curves is  simply the (total) time-delay between the
images.   Refsdal (1964)  proposed  to measure  time-delays in  lensed
supernovae, but his  method was published just when  the first quasars
were discovered (Schmidt 1963).  Quasars,  that later turned out to be
very numerous in the sky, rather bright, and photometrically variable,
were promising objects to measure time-delays if at least some of them
were  found  to  be lensed.   They  appeared  in  any case  much  more
promising  than  rare  and  transient phenomena  such  as  supernovae.
Indeed, thousands of  quasars are now known, and  several tens of them
are lensed.   Measured quasar time-delays  span over a broad  range of
values,  between  days  and  months.   One  is  larger  than  a  year:
Q~0957+561 (e.g., Vanderriest et al. 1989).

\begin{figure}[t]
\begin{center}
\leavevmode
\includegraphics[width=11.8cm]{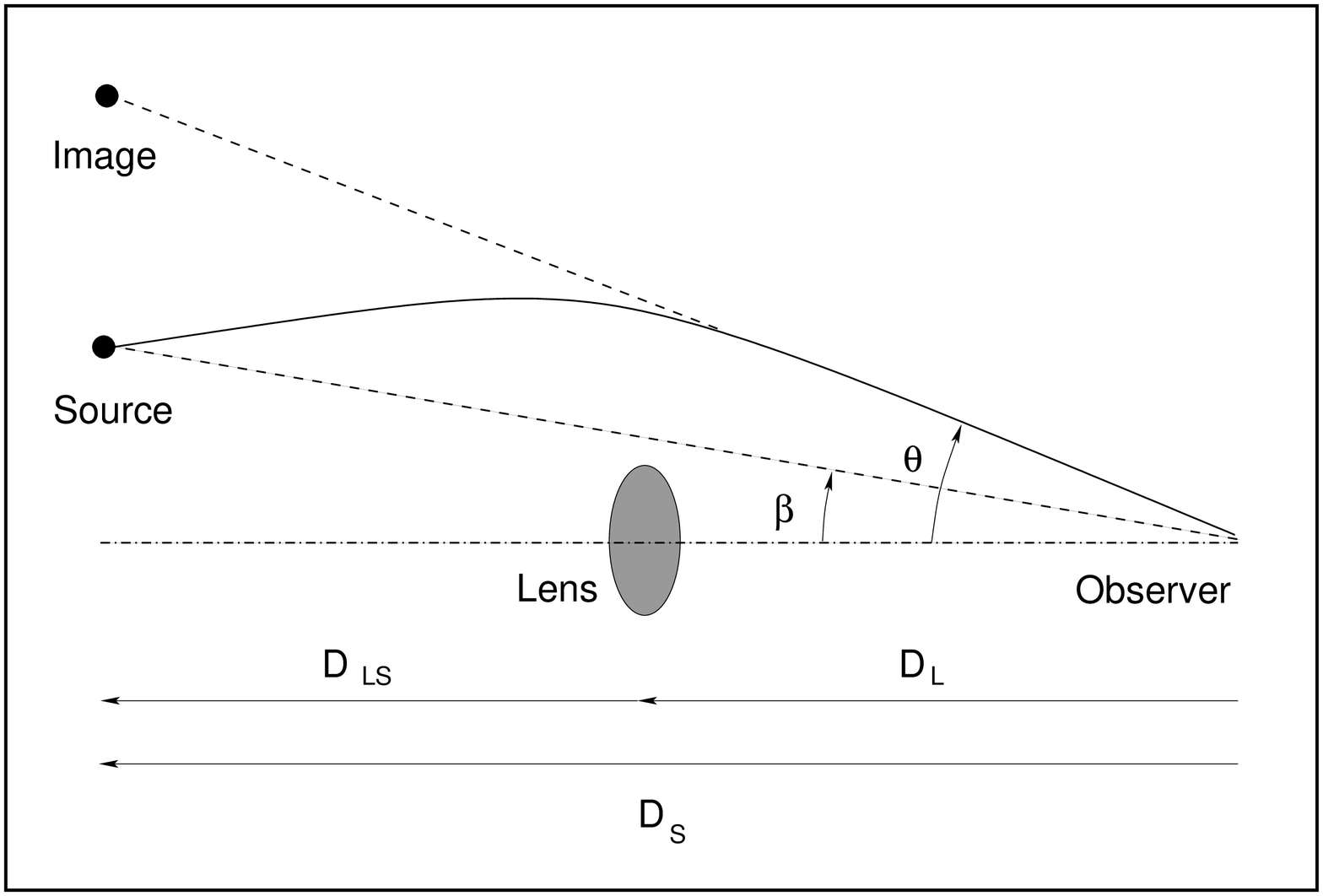}
\caption{Schematic  view  of  a  lensed  quasar, with  only  one  image
represented.  The  difference in length between  the straight (dashed)
and  curved (solid)  lines  is only  responsible  for the  geometrical
time-delay.  The  total time-delay also includes  a gravitational part
that  depends on  the mass  distribution  in the  lensing galaxy  (see
equations 1-3).}
\label{Lens_geom}
\end{center}
\end{figure}

\subsection{Constraints and Uncertainties}

Time-delays  can be  predicted from  lens modeling,  for  any observed
image configuration  and compared with  the measured ones in  order to
infer the  value of H$_0$.   The task requires  detailed observations,
deep, and  at high angular resolution,  and a good mass  model for the
lensing galaxy,  as can be seen  from the explicit  expression for the
time-delay in equations (1-3).   A full description of the calculation
can be  found in Schneider,  Ehlers \& Falco  (1992). We only  use the
result here to  illustrate how observations help to  achieve our goal.
As  explained  above,  the  total   time  delay  is  the  sum  of  two
contributions, so that:

\begin{equation}
t_{\rm tot} = t_{\rm geom} + t_{\rm grav},
\label{t-tot}
\end{equation}

Each contribution to the total time-delay writes as:

\begin{equation}
t_{\rm geom}(\vec{\theta}) = (1+z_{\rm L}) \frac{D_{\rm L} 
D_{\rm S}}{c D_{{\rm LS}}}(\vec{\theta}-\vec{\beta})^2,
\label{t-geom}
\end{equation}

\begin{equation}
t_{\rm grav}(\vec{\theta}) = (1+z_{\rm L}) \frac{8\pi G}{c^3}  \nabla^{-2} 
\Sigma (\vec{\theta}).
\label{t-grav}
\end{equation}

\vspace*{2mm}

where z$_L$ is the redshift  of the lensing galaxy.  As illustrated in
Fig.   3, the angle  $\vec{\theta}$ (in  2D in  real cases)  gives the
position of  the images  on the plane  of the  sky and $\beta$  is the
angular position of the source.

The Hubble parameter H$_0$ is contained in the geometrical part of the
time-delay, through the angular diameter distances to the lens D$_{\rm
L}$, to  the source D$_{\rm S}$  and through the  distance between the
lens and the source, D$_{\rm LS}$.

Equation (3),  the gravitational part of the  time-delay, depends only
on  well  known physical  constants,  and on  the  inverse  of the  2D
Laplacian of  the mass density  profile in the lensing  galaxy $\Sigma
(\vec{\theta})$. In other  words, it strongly depends on  the shape of
the 2D mass profile of  the lens (ellipticity, position angle), and on
its slope.  We  will see later that the main  source of uncertainty on
the gravitational  part of the  time-delay comes from the  {\it radial
slope} of the mass distribution.

Several of the ingredients necessary  to compute the time-delay can be
precisely measured from observations. Although every lensed system has
its own particularities, the positions of the lensed images defined by
$\vec{\theta}$ are  usually the easiest quantities  to constrain. With
present  day instrumentation,  an accuracy  of a  few  milli-arcsec is
reached. The  position of  the lensing galaxy  relative to  the quasar
images, when it  is not double or  multiple can be of the  order of 10
milli-arcsec. As for the position $\vec{\beta}$ of the source relative
to the  lens, it is  usually free in  lens models. No  observation can
constrain it.

In most  cases, astrometry  is not  a major limitation  to the  use of
lensed quasars. However, image  configurations that are very symmetric
about the center of the  lens are more sensitive to astrometric errors
than assymetric  configurations.  Let us assume  that $\vec{\beta}$ is
very small  compared with $\vec{\theta}$  (i.e., the source  is almost
aligned with the lens and  the observer).  Lets then consider a doubly
imaged quasar  with two  images located at  positions $\vec{\theta}_1$
and   $\vec{\theta}_2$  away   from   the  lens,   and  separated   by
$\vec{\theta}$.  The geometrical time-delay between the two images is:

\begin{eqnarray}
\Delta t_{\rm geom} = t(\vec{\theta_1}) - t(\vec{\theta}_2)   &  \simeq & \nonumber
(1+z_{\rm L}) \frac{D_{\rm L} D_{\rm S}}{c D_{\rm LS}}(\vec{\theta_1^2}-\vec{\theta_2^2}) \\
 & = & (1+z_{\rm L}) \frac{D_{\rm L} D_{\rm S}}{c D_{\rm LS}}
\biggr[ \vec{\theta}\cdot(2\vec{\theta}_1 - \vec{\theta})\biggr].
\label{t-geom-approx}
\end{eqnarray}

If  we   now  consider  that   the  error  on  the   image  separation
$\vec{\theta}$ is much smaller than the error on the position of image
1 relative to the lens, $\vec{\theta_1}$, we can approximate the error
on   the   time-delay.   Since   the   errors  $d\vec{\theta}_1$   and
$d\vec{\theta}$  on $\vec{\theta}_1$ and  $\vec{\theta}$ are  not much
correlated, they propagate on the time-delay as

\begin{equation}
\frac{d \Delta t_{\rm geom}}{\Delta t_{\rm geom}} 
\simeq \frac{2}{(|2\vec{\theta}_1 - \vec{\theta}|)}d\theta_1.
\label{t-geom-approx}
\end{equation}

In symmetric  configurations, where the  lens is almost  midway between
the  images, $2|\vec{\theta}_1|  \simeq |\vec{\theta}|$,  so  that the
denominator in equation  (5) is zero or close to  it, leading to large
relative errors on  the time-delay, and hence on  H$_0$, whatever mass
model is adopted for the lensing galaxy.

The advantage of symmetric  configurations over assymetric ones is that
they often have more than two lensed images (the source is more likely
to be  within the  area enclosed by  the radial caustic;  see previous
chapters on the basics of quasar lensing), offering the opportunity to
measure  several time-delays  per  system. The  drawback  is a  larger
sensitivity to astrometric errors.

Redshift information is also  of capital importance in the calculation
of the time-delay.  Time-delays are proportional to $(1+z_{\rm L})$ as
seen in equation  (4).  The angular diameter distances  also depend on
the redshift of the lens and source $z_{\rm L}$ and $z_{\rm S}$.  Both
should therefore be measured  carefully.  Although the lens and source
redshifts are available for most know system, their measurement is not
as straightforward  as one could  expect.  Given the  small separation
between the lensed images and the high luminosity contrast between the
source and lens, obtaining a spectrum of the lens is often challenging
and may involve significant struggling  with the data (e.g., Lidman et
al.  2000). In  other cases, for example in  systems discovered in the
radio, one  faces the opposite  situation: the optical  counterpart of
the source is  so faint that no spectrum can be  obtained of it, while
the lens is well visible (e.g., Rusin et al. 2001).

Finally,  the other cosmological  parameters such  as $\Omega_\Lambda$
and $\Omega_0$ also play a  role in the calculation of the time-delay,
through  the  angular  diameter  distances.   The  dependence  of  the
distances on  ($\Omega_\Lambda$,$\Omega_0$) is however  very weak.  In
addition, other methods (Supernovae,  CMB) seem much better at pinning
down their values  than quasar lensing does.  One  shall therefore use
the  known values of  ($\Omega_\Lambda$,$\Omega_0$) in  quasar lensing
and infer H$_0$, to which it is much more sensitive.

\subsection{The Mass Model and Degeneracies}

Most unknowns  in the  calculation of the  time-delay can  be measured
from deep  and sharp images (and  spectra), but we have  not paid much
attention, so  far, on the  gravitational part of the  time-delay.  It
depends   on   the   mass   surface   density   distribution   $\Sigma
(\vec{\theta})$ of  the lensing object(s),  which can not  be measured
directly.   It  has therefore  to  be  {\it  modeled}, and  getting  a
``realistic'' estimate of $\Sigma (\vec{\theta})$ is not trivial.

A  good  lens model  should  in principle  be  able  to reproduce  the
observables with  as few free  parameters as possible.   Ideally, this
model should be  unique.  It is ``asked'' to  reproduce the astrometry
of the quasar images with a  very high accuracy, as well as their flux
ratios.   In  fact  we  will  not  consider the  latter  as  a  strong
constrain.  Even  if flux ratios can  be measured with  an accuracy of
the order of the percent, they  are affected by extinction by dust and
by  microlensing events  due  to the  random  motion of  stars in  the
lensing galaxy (see Schechter  2003; Wambsganss 2003). Flux ratios may
therefore vary with time and  wavelength.  In addition, they should be
corrected for  the effect of  the time-delay, i.e., each  quasar image
should  be measured  when the  quasar  is seen  in the  same state  of
activity.  This  can be  done for quasars  with known  time-delays and
well measured light curves.  Curiously,  such a correction is not much
taken  into account  in the  literature, even  for quasars  with known
time-delays.

\begin{figure}[t!]
\begin{center}
\leavevmode
\includegraphics[width=11.7cm]{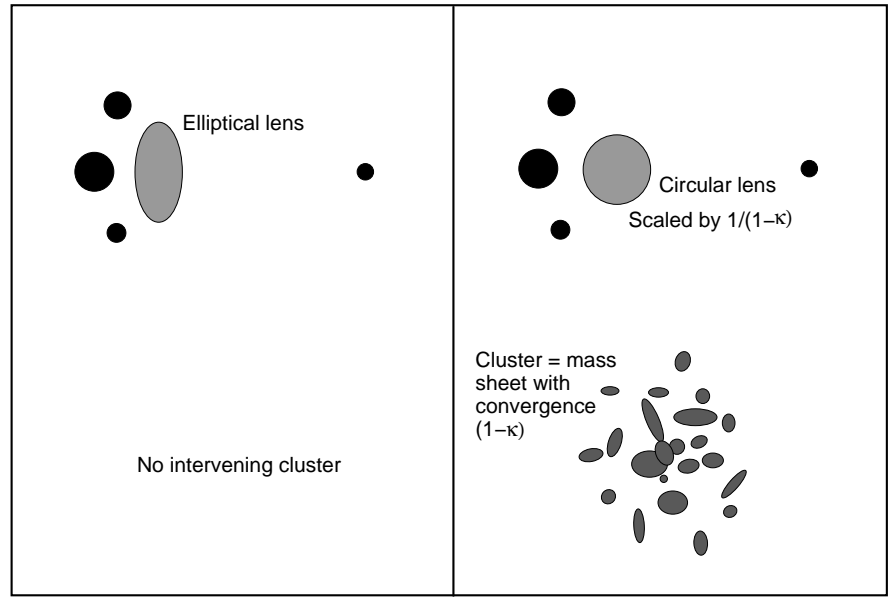}
\caption{Two ways  of obtaining a given image  configuration. The left
panel displays a system with four images, with an elliptical lens that
introduces convergence and shear at the position of the images. On the
right panel, is shown the same image geometry and flux ratios, but the
lens is  now circular. One would  in principle only  obtain two images
with  such  a  lens. The  shear  required  to  obtain four  images  is
introduced by the nearby cluster.   The mass density of the cluster is
represented through  its convergence $\kappa$.   The mass of  the main
lens  is  scaled  accordingly  by  1/(1-$\kappa$) so  that  the  image
configuration remains the  same as in the left panel:  the mass in the
main  lens and  in  the  cluster are  degenerate.   If no  independent
measurement is available for at least one of the components (main lens
or cluster), it  is often difficult to know,  from the modeling alone,
what exactly  are their respective  contributions.}
\label{Shear}
\end{center}
\end{figure}

When modeling  lensed quasars,  one is  asked, on the  basis of  a few
(usually 2  or 4)  quasar images, to  model the  whole two-dimensional
gravitational potential of the lensing galaxy or galaxies. There is of
course  no  unique solution  to  the  problem:  too few  observational
constraints are  available and several  mass models giving each  one a
different time-delay  can reproduce  a given image  configuration, its
astrometry  and flux  ratios. In  other  words, lens  models are  {\it
degenerate}.

Degeneracies  have  been  described   and  blamed  abundantly  in  the
literature for  being the  main source of  uncertainty in  lens models
(see for example Saha 2000;  Wucknitz 2002). Whatever precision on the
measured astrometry  and time-delay, several mass  models will predict
several  time-delays  and  hence   several  H$_0$.   One  must  devise
techniques to  break the  degeneracies or find  quasars that  are less
affected by them.

The main  degeneracy one has to  face in quasar lensing  is called the
{\it  mass sheet degeneracy}:  when adding  to a  given mass  model, a
sheet of  constant mass density (i.e.,  constant convergence $\kappa$,
as defined in the preceding chapters),  one does not change any of the
observables, except  for the time-delay.   The additional mass  can be
internal to the lensing galaxy  (e.g., ellipticity does not change the
total mass within the Einstein radius, but does change $\kappa$ at the
position of the  images) or due to intervening  objects along the line
of sight.  The  exact mass introduced by the  mass sheet increases the
total mass of the lens, but one can re-scale it and locally change its
slope at  the position  of the  images. The result  is that  the image
configuration does  not change, but  the convergence $\kappa$,  at the
position  of the  images  does change,  and  modifies the  time-delay.
Therefore, knowledge of the the {\it slope} of the mass profile of the
lensing  galaxy, whether  it be  under the  form of  a model  or  of a
measurement, is  one of  the keys to  the determination of  a ``good''
model.

Changing the slope of the lens will change $\kappa$ at the position of
the images, but adding intervening  objects along the line of sight to
the  lens has  a  similar effect.   A  group or  cluster of  galaxies,
located angularly close to the  lens, will add its own contribution to
the  total  mass  density at  the  position  of  the images.   If  the
group/cluster  has a  constant density  $\kappa$, rescaling  the total
mass of the  lensing galaxy by $1/(1-\kappa)$ will  leave the observed
images configuration  unchanged, as illustrated  in Fig.  4,  but will
change the time-delay.

Adding   convergence  also modifies  the    shear $\gamma$, hence  the
ellipticity of the main  lens.   There are  therefore several ways  of
reproducing a  given combination   of  shear  and convergence at   the
position of the images, as illustrated in Fig. 4. In the left panel of
the  figure, the shear  $\gamma$,  is produced only  by  the main {\it
elliptical} lensing galaxy.  In the right panel,  the total shear is a
combination of the lens-induced shear and of that of the nearby galaxy
cluster. In  principle, it is  even possible to   model a given system
equally with  either one single   elliptical lens or with a  completely
circular   lens and    an   intervening cluster   responsible  for  an
``external'' source of shear.

Both types  of degeneracies can be  broken or, at least,  their effect
can  be strongly minimized, by  constraining in an independent way (1)
the mass  profile  of the   main  lens, and  (2)  the  total mass (and
possibly  also  the radial mass   profile) of any  intervening cluster
along the line of sight.  This work can be done with detailed imaging,
spectroscopy of all  objects  along the line  of  sight, and  by using
numerical  multi-components  models  for  the total lensing potential.
This is the topic of the next section, illustrated through the example
of the well studied quadruple PG~1115+080.

\section{PG~1115+080: a Clean Quadruple}

\pg\ is one  of the first case of quadruply  imaged quasar (Weymann et
al. 1980) and among the best studied  lensed quasars, one of the still
rare  systems    with  well   measured    time-delays  (Schechter   et
al. 1997). The  tremendous  gain in spatial resolution  achieved since
the eighties (see  Fig.~5), has allowed  detailed understanding of the
system.  Since  the range of observational data  available for \pg\ is
so broad,  we take this object as  an example to show how observations
help to pin down the Hubble parameter H$_0$.

\begin{figure}[t!]
\begin{center}
\includegraphics[height=5.7cm]{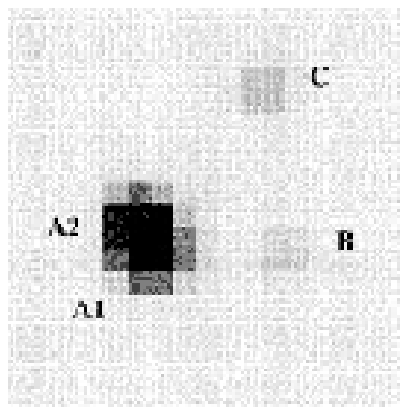}
\includegraphics[height=5.7cm]{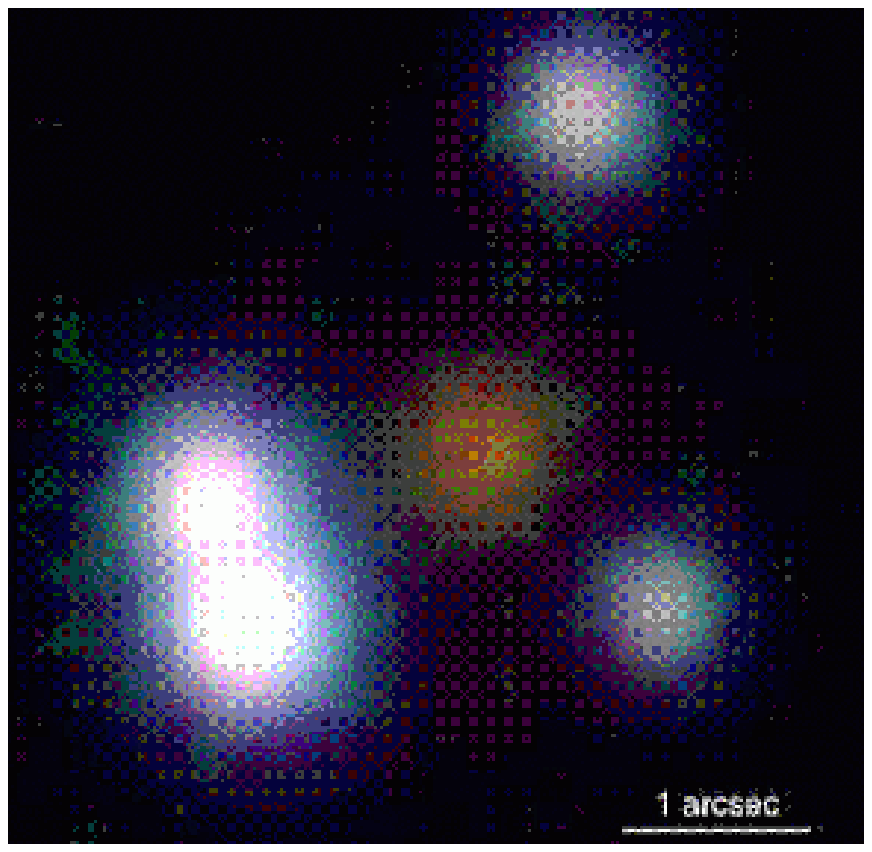}
\caption{{\it Left}: ground based image of \pg\ adapted from Schechter
et al (1997), with a field of  view of 5\arcsec\ on a side.  The data,
with a resolution of about 0.8\arcsec, is of sufficiently good quality
for measuring the  photometric variations of the quasar  images. It is
however not good  enough to obtain detailed surface  photometry of the
lensing galaxy. {\it Right:} one of the very best images ever obtained
of  \pg, with  the  8.2m Subaru  telescope.   The  image has  a
resolution  of 0.32\arcsec,  approaching  that  of the  Hubble
Space  Telescope  (Iwamuro et  al.  2000),  and  allowing for  precise
measurements of  the quasar  and lens astrometry  as well  as for detailed
surface photometry of the lens. }
\label{pg1115_demo}
\end{center}
\end{figure}

\pg\ is a bright quasar at z=1.722, with a relatively ``wide'' angular
separation between  its images, of the order  of 2\arcsec.  Two images
are isolated (B and C in Fig.  5) and  a blend of two brighter images,
is located on the other side of the lens in projection on the plane of the sky
(A1/A2 on Fig. 5).  This blend A1/A2 was not resolved on the discovery
images of the object,  and  \pg\ was   subsequently believed to  be  a
triple  until  more detailed observations revealed  it  was  in fact a
quadruple   (Young et al. 1981). Applying   Refsdal's method (1964) to
\pg\ has been  a long process.   Some of the  ingredients necessary to
the  lens modeling had to wait  for years before suitable observations
finally became available.

\subsection{Redshifts}

\begin{figure}[t!]
\begin{center}
\includegraphics[height=9cm]{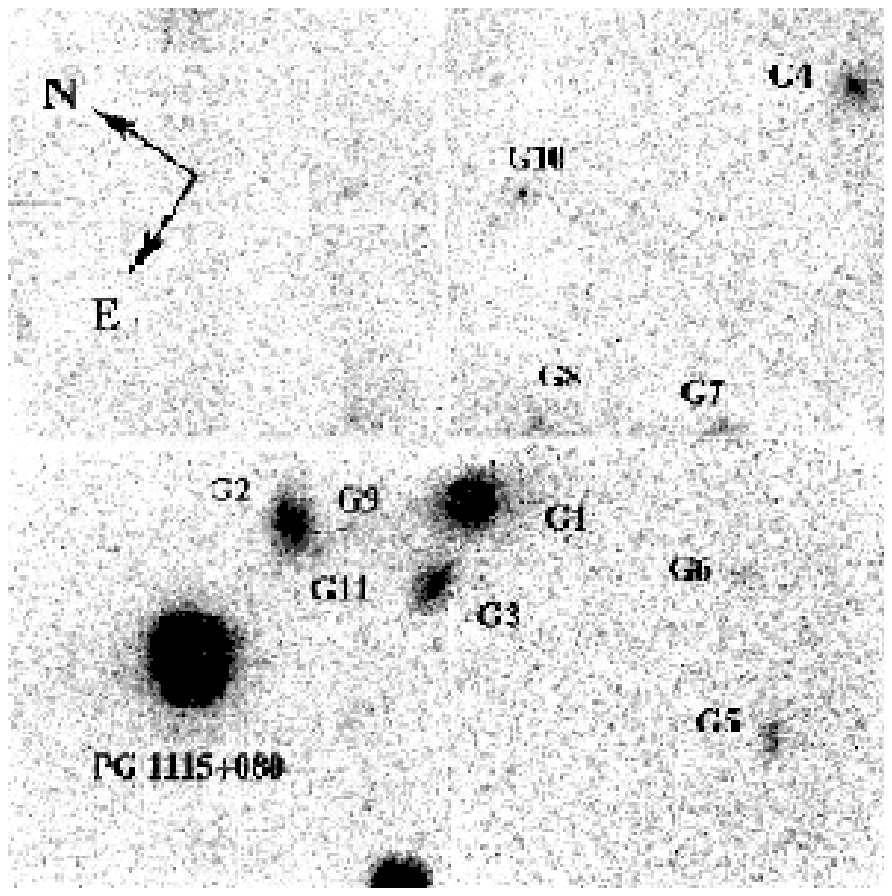}
\caption{Infrared  image of  \pg, taken  with the  HST and  the NICMOS
camera (Impey et al. 1998). The field of view  is 35\arcsec\ on a side
and shows  several ``companions'', projected on  the plane of the sky.
The galaxies labeled ``G'' are part  of a group  to which also belongs
the main lensing  galaxy  \pg. These galaxies  have  to be taken  into
account  when  modeling the  potential well  responsible for  the image
configuration.}
\label{pg1115_cluster}
\end{center}
\end{figure}

As already mentioned, the redshift  of the lensing galaxy $z_{\rm L}$
is mandatory for the  determination  of any  model. Although \pg\  was
discovered in 1980, it is  only in 1997 that  $z_{\rm L}$ was measured
(Kundic et al.  1997), by using the  Keck telescope.  The galaxy is at
low redshift, z$_{\rm L}=0.311$.  With  deeper exposures, its velocity
dispersion  has also  been measured,  $\sigma_{\rm  L}  = 281  \pm 25$
km.s$^{-1}$ as well as the velocity dispersion  of a group of galaxies
found along the line  of  sight (Fig.   6), $\sigma_{\rm grp}  =  326$
km.s$^{-1}$   (Tonry 1998).  These   measurements  provide us with  an
independent estimate of  the total mass  of  the lens and  intervening
group.     They are crucial  when  trying    to break  the mass  sheet
degeneracy.

\subsection{Time-delays and Temporal Sampling}

With  its bright  images, its  wide angular  separation, and  with the
availability of reference  stars within a few arc-minutes,  \pg\ is an
excellent  target  for  photometric  monitoring.  Excellent  does  not
automatically imply easy !  Photometric monitoring still requires good
seeing, at least a medium  size telescope (2m) and a temporal sampling
adapted to the observed variations. 

Choosing the temporal sampling of  the light curves is critical. It is
often  claimed  that it  should  be  smaller  than half  the  expected
time-delay,  but this  is  not  quite true.   The  measurement of  the
time-delay  is  done  through   the  measurement  of  the  photometric
variations  of the quasar  itself. Therefore,  the sampling  should be
chosen according to the typical time-scale observed or expected in the
photometric  light  curves.  In  an  ideal  case  where the  intrinsic
variations of  the source are  slow compared with the  time-delay, one
can in  principle chose  a sampling even  larger than  the time-delay.
The situation is quite the same than in image processing where one can
measure the  position of an object on  a CCD frame with  a much higher
accuracy than the sampling adopted to represent this signal.

\begin{figure}[t!]
\begin{center}
\includegraphics[width=12cm]{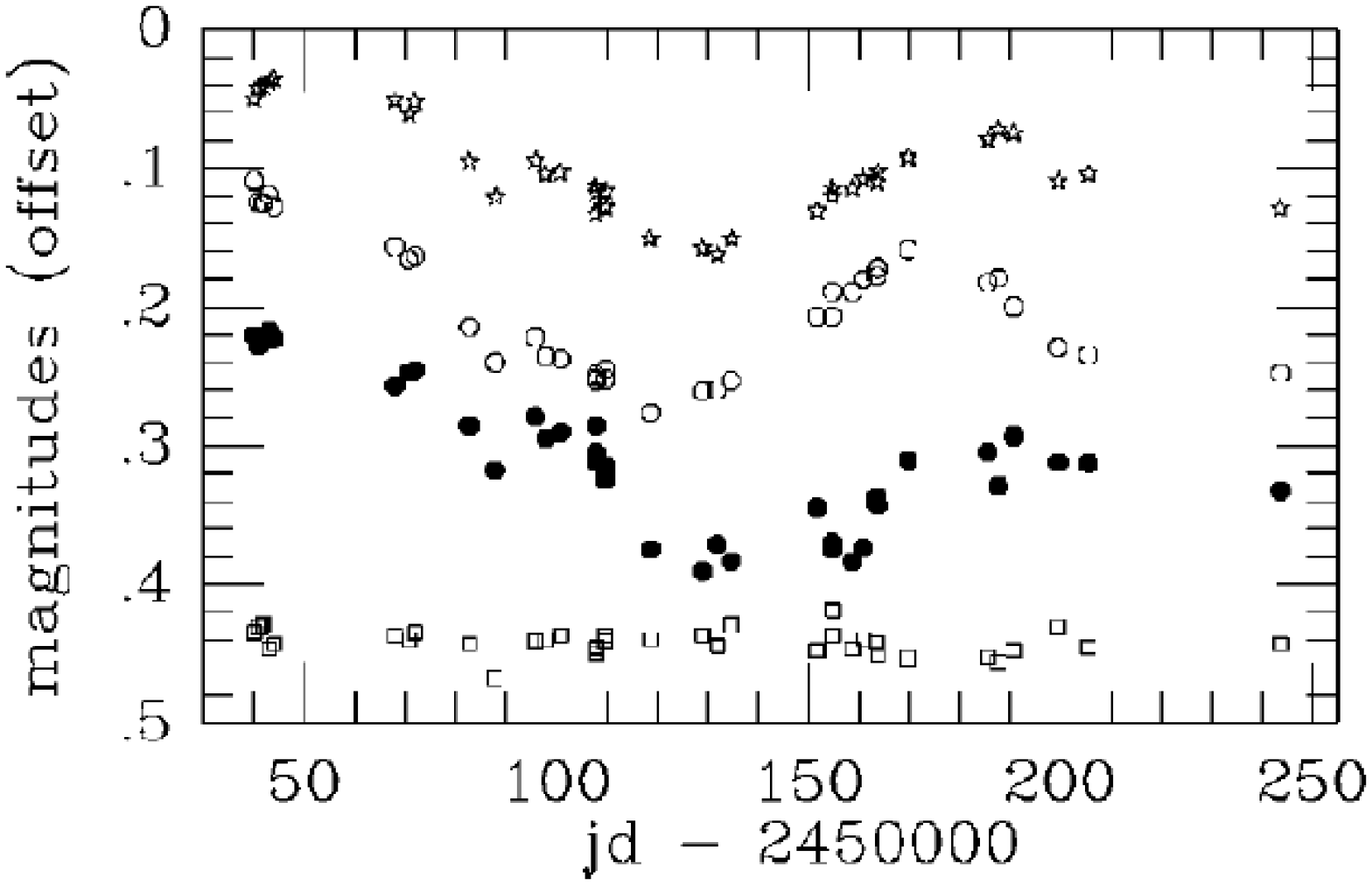}
\caption{Optical $V$-band light curves for the different components of
\pg. From top to bottom, the  curves for the blend A1/A2, component C,
component B, and  a reference star (Schechter et  al.  1997). Note the
smooth and slow variation of the quasar, the absence of high frequency
flickering,  as   would  introduce  microlensing,   and  the  striking
similarity between the light curves.}
\label{pg1115_lc}
\end{center}
\end{figure}

\begin{figure}[p]
\begin{center}
\includegraphics[width=10cm]{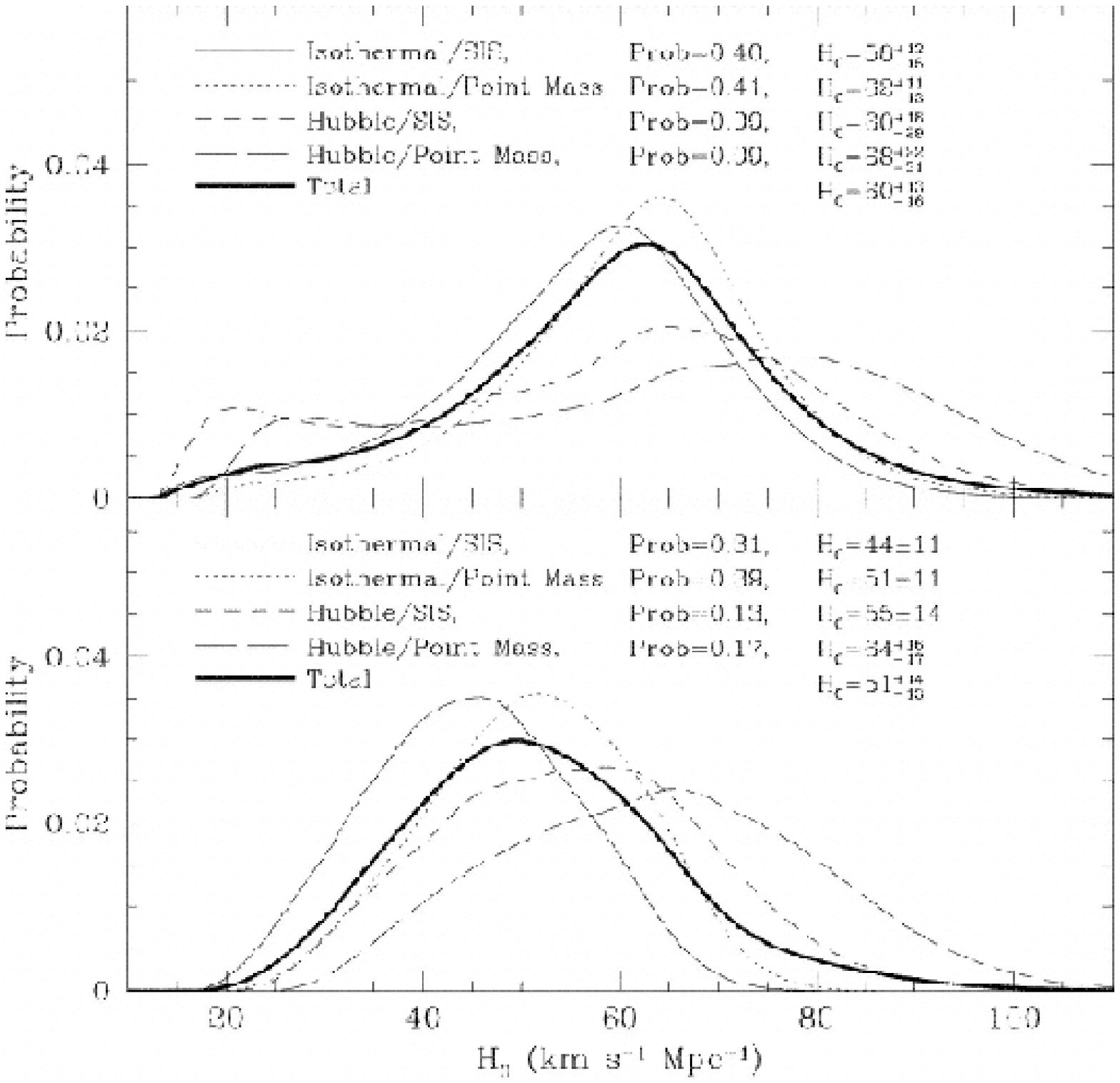}
\includegraphics[width=10.5cm]{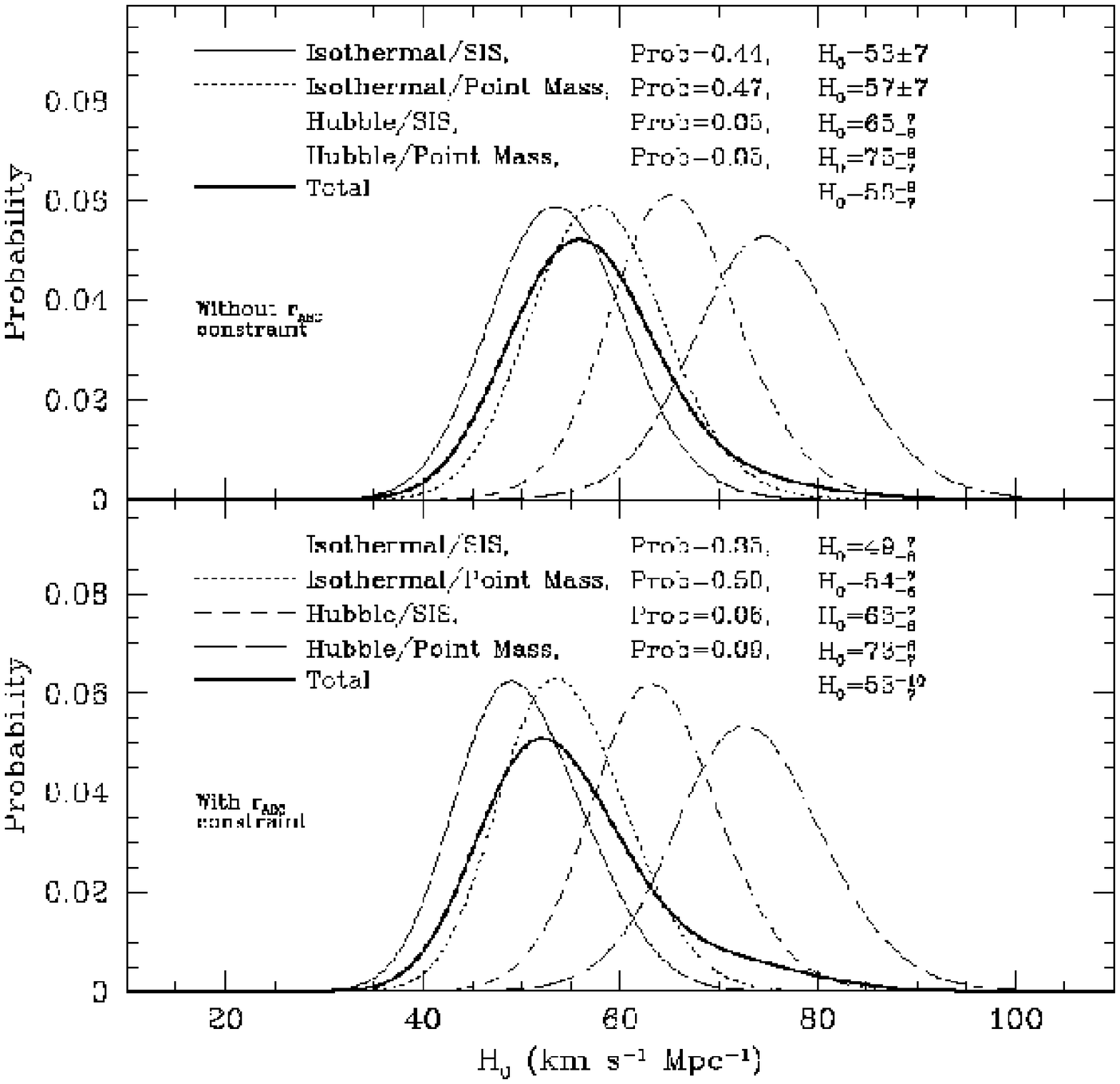}
\caption{{\it Top:} probability density  for H$_0$ using the models of
Keeton \& kochanek  (1997), for various combinations of  models for the
main lens and intervening group  and using one time-delay (top panel)
or two  time-delays (bottom  panel). {\it Bottom:}  Same as  above but
using improved astrometry (Courbin et al. 1997).}
\label{pg1115_keeton}
\end{center}
\end{figure}

However, while   one do  not need  in  principle   very high  temporal
sampling, the situation in  practice is just  the opposite. First, one
does not know  in advance what the time-scale  of  the variations will
be: the frequency range in quasar variability spans from days to weeks
or even  months.   Second, even  if  the  variations are  slow,  their
amplitude is small, often of  the order of  a few tens of a magnitude,
sometimes  a bit  more in the   case of  Broad  Absorption Lines (BAL)
quasars. Measuring faint variations is  easier with well sampled light
curves. Well sampled curved  are, in addition,  well suited to the use
of cross-correlation techniques  in  order to measure the  time-delay.
Finally,   microlensing  by stars   in   the lensing galaxy introduces
flickering of  the light curves, with a  frequency that is  in general
unknown and not   even easy to predict before   the data are  actually
taken.  There is therefore no general  line to adopt  on the choice of
the sampling other than trying to get ``the finest possible sampling''
!

The case of  \pg\ is lucky enough that  the photometric variations  of
the source are slow,  and well sampled light  curves could be obtained
(Schechter et al.  1997; see Fig.  7). Even with light curve of such a
quality, extracting the time-delay(s) can  be tricky. Schechter et al.
(1997)  uses the method  devised by Press  et al.  (1982) to compute a
global  $\chi^2$ between shifted  versions of  the light curves.  They
estimate the time-delay between components C and B to be $\Delta t(CB)
= 23.7 \pm 3.4$ days and  the time-delay between  component C and A is
$\Delta t(CA)  = 9.4$ days, where  C is always  the first one to vary,
i.e.,  the  ``leading image''.   A  different  approach was  chosen by
Barkana  (1997)  to  analyse  the  same  data.    Using an  analytical
representation   of the  light curves  and    taking into  account the
correlations  between  the   errors  on  the   individual  photometric
measurements, they derive a very similar time-delay between components
C and  B, $\Delta t(CB) =  25^{+3.3}_{-3.8}$ but the second time-delay
is significantly longer $\Delta t(CA) = 13$ days.

\subsection{Modeling and Influence of the Astrometry}

Using a  simple isothermal model  plus external shear (induced  by the
cluster) Schechter et  al.  (1997) infered H$_0$ =  42 \kmsmpc, but no
detailed  observations of  the lensing  galaxy was  available  at that
time. In fact,  Schechter et al. (1997) found that a  value as high as
84  was possible  as  well.  Keeton  \&  Kochanek (1997)  investigated
analytical  models  for   the  lensing  galaxy,  including  isothermal
profiles and softened power laws with core radius. They found that the
potential well  in \pg\  could not be  modeled using a  single lensing
galaxy, whatever its mass profile.  External shear by the nearby group
had to  be invoked. A combination  of models were tested  for the main
lens and for  the cluster, and analyzed in a  statistical way in order
to derive a  probability density for H$_0$. The  experiment is done in
two ways:  by taking into account  only one time-delay in  \pg, and by
using the two measured time-delays.

The models by Keeton \& Kochanek are based on astrometry obtained with
HST data,  before the optics  was refurbished (Kristian et  al. 1993),
and with  rather large astrometric errors.  As  systems with symmetric
image configurations  about the lens  are more sensitive to  errors on
the astrometry than assymetric ones,  the result by Keeton \& Kochanek
could  be improved further,  simply by  improving the  HST astrometry.
This  was done by  applying deconvolution  techniques to  ground based
images of \pg. Fig.~8 illustrates  the gain in the accuracy.  With the
new astrometry, not only the width of the probability distributions is
decreased,  but there  is  also better  agreement  between the  curves
obtained for  one and for  two time-delays.  The  combined probability
for the pre-refurbishment HST  astrometry is H$_0$ of 51$^{+14}_{-10}$
\kmsmpc  while  H$_0$  =  53$^{+10}_{-7}$ \kmsmpc  with  the  improved
astrometry.

In parallel with the efforts  to improve the observational constraints
for \pg, new models were developed, with the aim of exploring more of
the  parameter space defined  by the slope of the  mass profile of the
lens, its   ellipticity and its position  angle.   One approach to the
problem is  to consider a family of  non-parametric models, where the
lensing galaxy is decomposed on a  grid of mass ``pixels'' or tiles.
The   astrometry,   flux   ratios, and     time-delays are  given   as
observational  constraints,  and the  output of   the procedure  is  a
pixelized mass map of the lens, as well as its slope.   As is done for
parametric  models, one can  have  a statistical  approach  and run many
models in order to infer a probability density  for H$_0$ (Williams \&
Saha 2000). As the  range of possible models is  much broader with the
non-parametric  approach  than with the  parametric one,  it was found
that H$_0$ could be anything between 42 and 84, given the observations
available for \pg\ at that time (Saha \& Williams 1997) !

\begin{figure}[t!]
\begin{center}
\includegraphics[width=11cm]{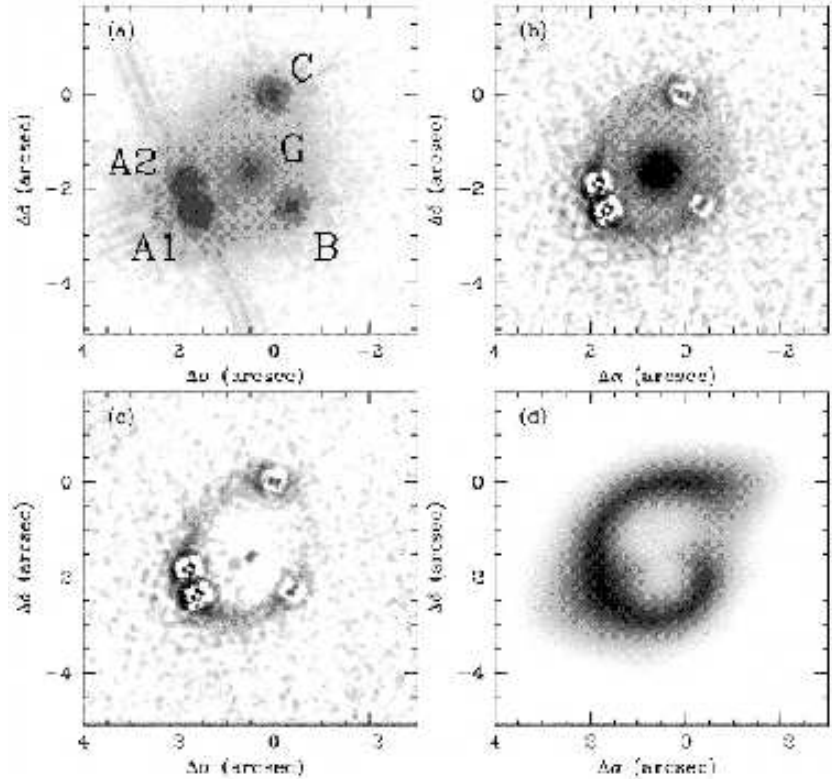}
\caption{An HST image of \pg\ is  shown in the upper left panel (Impey
et al. 1998), with the four quasar images and the lens G. Removal of the
quasar images, as done in the upper right panel, unveils an almost full
ring: the  lensed host galaxy of  the quasar. It is  even more evident
after removal  of the lensing galaxy,  in the lower left  panel.  If a
model is  adopted for the ``unlensed''  host galaxy, the  shape of the
lensed ring  can be predicted  (lower right panel), and  compared with
the observations.}
\label{pg1115_ring}
\end{center}
\end{figure}

\subsection{HST Imaging: the Quasar Host Galaxy}

Improving the astrometry over and over, helps,  but does not solve the
fundamental problem imposed by the  mass sheet degeneracy.  The number
of lensed images available  can not  be increased, so that potential
well of the lens, and  hence the arrival-time  surface, are probed only
at a very limited number of points.  This is quite not true as soon as
one considers  extended sources. Quasars have  host galaxies. They are
faint, but they have a much  larger angular size  than the quasar they
harbourgh.  The  effect of   lensing  on their shape,   distorted  and
stretched, is much stronger than on the  point source quasar.  The
role  of high  angular  resolution observations therefore  extends well
beyond  the  simple   aim   of measuring  the    image  positions  and
fluxes. Each detail discovered in the lensed  image of the host galaxy
is an additional point in the mass map.

\begin{figure}[t!]
\begin{center}
\leavevmode
\includegraphics[height=5.9cm]{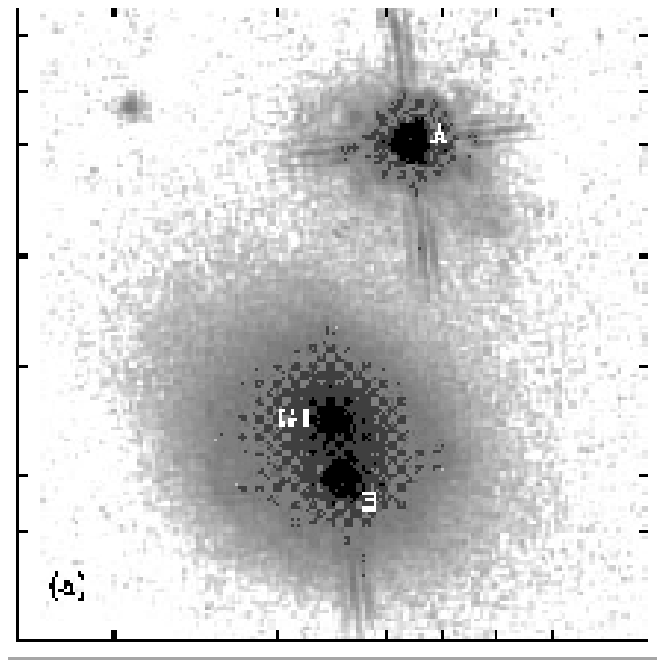}
\includegraphics[height=5.8cm]{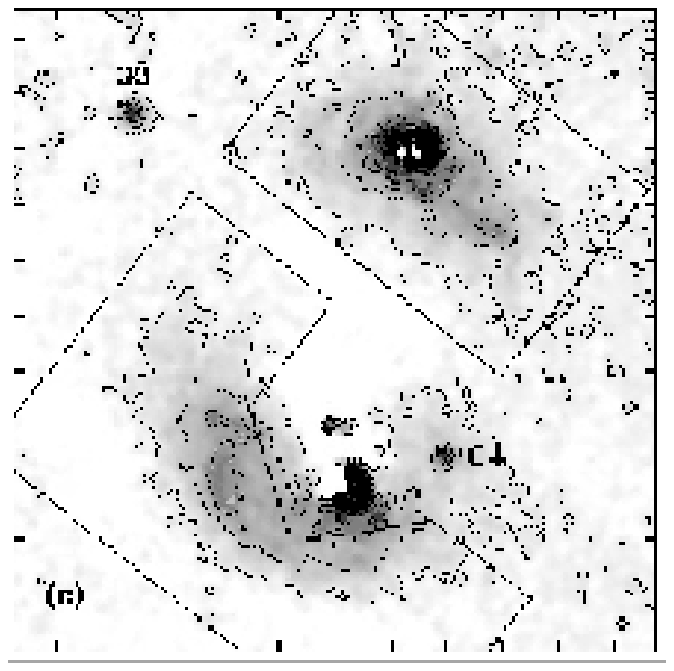}
\caption{{\it  Left:} HST  NICMOS observation  obtained by  the CASTLE
group,  where  the  two  quasars  and  the  lensing  galaxy  are  well
separated.  {\it Right:} the quasars and the lens have been removed to
unveil the  host galaxy  of Q~0957+561. On  the contrary of  \pg, many
details are seen in each image of the host (Keeton et al.  2000).}
\vspace*{2mm}
\leavevmode
\includegraphics[height=5.9cm]{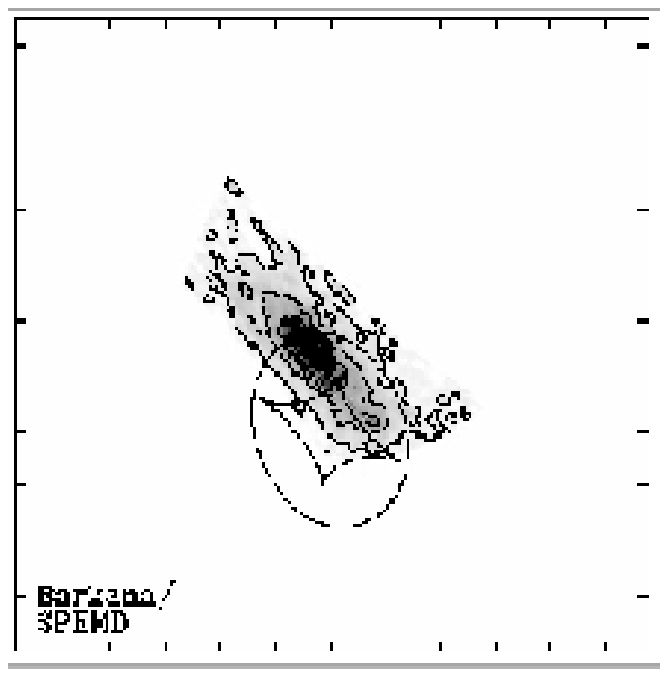}
\includegraphics[height=5.8cm]{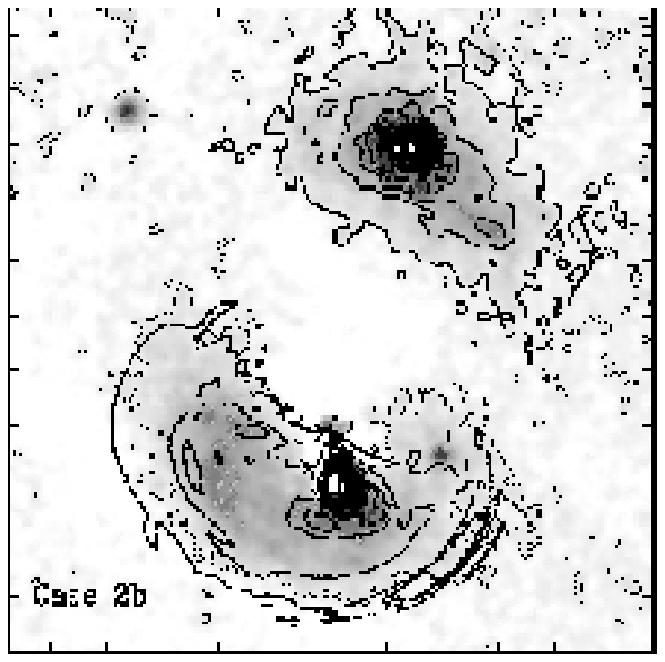}
\caption{{\it Left:} the  A image of the host  is mapped
back into  the source plane to  reconstruct the unlensed  image of the
host.  {\it  Right:} the  source is imaged  again in order  to produce
image B given  the lens model. The image B, as  predicted by the model,
is the solid  contour in the figure. Grey  levels are the observations
(Keeton et al.  2000).}
\label{Q0957_keeton}
\end{center}
\end{figure}

\pg\  has been  observed with  the  refurbished HST,  in the  infrared
(Impey et al. 1998), with  a resolution and depth sufficient to unveil
an almost full ring joining the quasar images (Fig. 9). While the host
galaxy provides  more constraints  for the models,  it also  forces to
introduces more  degrees of  freedom. Indeed, a  model has to  be {\it
chosen} to represent the ``unlensed'' host, for example an exponential
disk, as done  in Impey et al. (1998). Feeding  the lensing model with
the  image  positions,  time-delays,  flux ratios,  lens  ellipticity,
position of the nearby group of  galaxies, and the model for the host,
one can reproduce the observed ring.

The actual  gain  brought by the  ring  is significant  but not always
sufficient to discriminate between lens  models.  Impey et al.  (1998)
give two  estimates for H$_0$ for two  different type  of lens models.
One assumes that light traces mass, i.e., that the mass-to-light ratio
is constant  in  every point of the  lens.  The second, more realistic
according   to  our current   knowledge   of the  mass distribution in
galaxies, includes  a  dark halo  component.  Using  these two models,
they find H$_0$ = 66 \kmsmpc and H$_0$ = 44 \kmsmpc, respectively.

A complete  Einstein  ring was  found  in  \pg, but  with no  resolved
details, only the   surface brightness can  be used   to constrain the
models. Because  the of the high degree  of symmetry of the system, and
given the resolution of the observations, it  is not clear whether full
rings  are actually useful  for   sorting out  between different   mass
models, or not (see, e.g., Saha \& Williams, 2001).

Many other lensed hosts galaxies have been discovered using HST images
(Lehar et al.  2000), in systems that are less symmetric than \pg, and
with higher degree of detail.  A good example is the double Q~0957+561
in which the host galaxy  reveals clumps, distorted arcs and for which
a different approach of the modeling was used, not involving any prior
model for the host's surface  brightness (Keeton et al.  2000).  Since
the  observations of  the host  of  Q~0957+561 are  probably the  most
detailed existing  so far,  we will briefly  leave \pg\ aside  and use
Q~0957+561  to illustrate  the use  of hosts  in quasar  lensing.  The
observations  are  shown in  Fig.~10,  while  Fig.~11 illustrates  the
modeling method. Based on the image of the brighter component A of the
host,  mapping it back  onto the  source plane,  the unlensed  host is
reconstructed. It is  then re-imaged, in order to  produce the fainter
image B, and  a $\chi^2$ is minimized, between  the observed image and
the prediction from the model.  The method takes advantage of the high
degree of details in the  host, and on the assymetric configuration of
the system.   It does  not require any  strong prior knowledge  on the
host.   The  method  allows  to  rule out  several  models  previously
published in the literature.

\subsection{Yet Further Constraints: Spectroscopy}

So  far,  and  mainly   for  technical  reasons,  the  most  important
constraints on  lens models were coming from  imaging. Spectroscopy is
nevertheless  another  key  component,  that will  become  of  growing
importance in the near future,  thanks to 3D spectrographs mounted on
large telescopes.

Spectroscopy provides independent access   to  the total mass of   the
lenses: whenever an intervening cluster or group  is seen close to the
line of  sight to a lens,  measuring its velocity  dispersion helps to
break the mass sheet degeneracy it introduces with the mass of the main
lens (e.g., Falco et al.  1997; Tonry 1998; Kneib et al.  2000).  If a
measurement of  the velocity dispersion  of the main lens is available
as well, the effect of the mass sheet  degeneracy is greatly minimized
In the case of \pg, both the velocity  dispersion of the main lens and
nearby group are available (see Table 1).

For  an  isothermal  sphere  the  line of  sight  velocity  dispersion
$\sigma$ is directly related to the Einstein radius, through the H$_0$
independent relation: 

\begin{equation}
\theta_{\rm E} = 4\pi \frac{\sigma^2}{c^2}\frac{D_{\rm L}}{D_{\rm S}}
\label{veldisp}
\end{equation}

Since the Einstein  radius $\theta_{\rm E}$ only depends  on the total
mass of the lens (within  the Einstein radius), measuring the velocity
dispersion does not yield the  mass profile of the lens. Nevertheless,
it can  be used to put  strong constraints on the  mass profile, using
further knowledge  we have  on galaxies in  general. Treu  \& Koopmans
(2002) take  the problem  under  this view  angle  and feed  dynamical
models of galaxies  with the measured velocity dispersions  in \pg. In
their model, the lensing galaxy is composed of a dark matter halo with
a  mass   density  profile   of  the  form   $\rho  (r)=r^{-{\gamma}}$
($\gamma=1$ for a Navarro, Frenk  \& White (1997) profile) and a Jaffe
stellar  density profile (Jaffe  1983). New  parameters appear  in the
model: the slope $\gamma$ or the dark matter halo, and the fraction of
the galaxy  that is under  the form of (luminous)  stars, f$_{\star}$.
However,  ($\gamma$,  f$_{\star}$) define  a  parameter  space with  a
rather sharp peak, centered at  $\gamma \sim 2.35$ and f$_{\star} \sim
0.67$, for \pg. There is therefore a rather small range of lens models
that fit simultaneously  the physical properties of the  lens, and the
constraints imposed by  lensing.  Using this approach leads  to H$_0 =
59^{+12}_{-7}$  \kmsmpc,  a  value  which  is much  less  affected  by
systematic  errors  than other  estimates,  not  taking the  dynamical
information into account.

\renewcommand{\arraystretch}{1.3}
\begin{table}[p!]
\caption{Summary  of   the  observational  saga  of   \pg\  since  the
time-delay measurement, and estimates  for H$_0$ (along with the 1$\sigma$
errors, when available) while the observations improve.}
\vspace*{2mm}
\begin{tabular}{llc}
\hline
Reference                & Observational or           &  H$_0$  \\ 
                         & theoretical improvement    & (\kmsmpc) \\
\hline \hline
Schechter et al. (1997)  & Time delay measurement:     &   \\
                         & $\Delta(CB)=23.7 \pm 3.4$ days  &  42 or 84  \\
                         & $\Delta(CA)=9.4$ days         &            \\
\hline
Barkana (1997)           & Redetermination of            & \\
                         & time-delays (Schechter's data): & no new \\
                         & $\Delta(CB)=25^{+3.3}_{-3.8}$ days  &  estimate \\
                         & $\Delta(CA)=13$ days         &  \\
\hline
Kundic et al. (1997)     & Redshifts of the lens            & \\
                         & and group:                        & 52$\pm$14\\
                         & z$_{\rm lens}$ = z$_{\rm group}$ = 0.311      & \\
                         & $\sigma_{\rm group}$ = 270 $\pm$ 70 km/s & \\ 
\hline
Saha \& Williams (1997)  & Non parametric models including   &  \\
                         & main lens plus group. Explore     & \\
                         & broad range of lens mass profiles & 42, 63, or 84    \\
                         & and ellipticities.                 & \\
\hline
Keeton \& Kochanek (1997)& Parametric models including    &   \\
                         & main lens and group. Explore   & 51$^{+14}_{-13}$ \\
                         & broad range of lens shapes.    &  \\
\hline
Courbin et al. (1997)    & Improved astrometry from  &   \\
                         & ground based imaging.      & 53$^{+10}_{-7}$ \\
\hline
Impey et al. (1998)      & New HST/NICMOS images. &   \\
                         & Discover lensed ring.  & 44 $\pm$ 4 or 65 $\pm$ 5  \\
                         & Improved astrometry and lens & \\
                          & ellipticity.                   & \\
\hline
Tonry (1998)             & Velocity dispersion of         &   \\ 
                         & the lens and group:             &   \\
                         & $\sigma_{\rm lens}=281\pm25$ km/s &   \\
                         & $\sigma_{\rm group}=326$ km/s  &   \\
\hline
Treu \& Koopmans (2002)  & Use dynamical info on the      & \\
                         & lens and group to break        & 59$^{+12}_{-7}$ ($\pm$3 syst) \\
                         & degeneracies between models    &  \\
\hline
\end{tabular}
\end{table}

\renewcommand{\arraystretch}{1.3}
\begin{table}[p!]
\caption{Summary of  measured time-delays,  given with respect  to the
leading image.  When several  measurements have been obtained, we only
give the more recent one.  The 1$\sigma$ error as well as the relative
error on the time-delay are given. The third column displays the value
infered for  H$_0$ for each system  and the 1$\sigma$  error.  In most
cases  we have  considered  the value  given  in the  paper where  the
time-delay  was  published. When  H$_0$  has  been  obtained by  other
authors, the  reference is given as  well.}
\vspace*{2mm}
\begin{tabular}{llc}
\hline
Object                   & Time-delay(s)                    &  H$_0$    \\ 
                         & (leading image first)            & (\kmsmpc) \\
\hline \hline
B~0218+357               &                                          &                   \\
Biggs et al. (1999)      & $\Delta(BA)$ = 10.5$\pm$0.2 days (2$\%$) & 69$^{+7}_{-9}$  \\
\hline
RX~J0911+0551            &                                          &                        \\
Hjorth et al. (2002)     & $\Delta(BA)$ = 146$\pm$4 days  (3$\%$)   & 71$\pm$2 ($\pm$4 syst) \\
\hline
Q~0957+561               &                                              &   \\
Kundic et al. (1997)     & $\Delta(BA)$ = 417$\pm$0.6 days  (0.15$\%$)  & 64$\pm$7 \\
                         & but multiple time-delays                     & \\
Goicoechea  (2002)       & $\Delta(BA)$ = 425$\pm$4.0 days  (1.0$\%$)   & \\
                         & $\Delta(BA)$ = 432$\pm$1.9 days  (0.5$\%$)   & \\
\hline 
HE~1104$-$1805           & Strong micro/milli-lensing                   &           \\
Gil-Merino et al. (2002) & $\Delta(AB)$ = 310$\pm$9 days  ($3\%$)       & 48$\pm$2 or 62$\pm$2 \\
\hline
PG~1115+080              &                                              &           \\
Schechter et al. (1997)  & $\Delta(CB)=25^{+3.3}_{-3.8}$ days  (14$\%$) &           \\
Barkana (1997)           & $\Delta(CA)=13$ days                         &           \\
Treu \& Koopmans (2002)  &                                              & 59$^{+12}_{-7}$ ($\pm$3 syst) \\
\hline
SBS~1520+53              &                                              &           \\
Burud et al. (2002)      & $\Delta(BA)$ = 130$\pm$3 days (3$\%$)        & 51$\pm$9  \\
\hline
B~1600+434               & Radio and optical time-delays          &                \\ 
Koopmans et al. (2000)   & $\Delta(BA)$ = 47$\pm$5 days (10$\%$)  &  57$^{+7}_{-6}$ \\
Burud et al. (2000)      & $\Delta(BA)$ = 51$\pm$2 days (4$\%$)   &  52$^{+7}_{-4}$ \\
\hline
B~1608+656               &                            &           \\
Fassnacht et al. (2002)  &  $\Delta(BA)$ = 31.5$\pm$2 days (6$\%$)  & 63$\pm$15 \\
                         &  $\Delta(BC)$ = 36.0$\pm$2 days (6$\%$)  &           \\
                         &  $\Delta(BD)$ = 77.0$\pm$3 days (4$\%$)  &           \\
\hline
PKS~1830$-$211           &                                              &  No estimate. Lens \\
Lovell et al. (1998)     & $\Delta(BA)$ = 26$^{+4}_{-5}$ days  (17$\%$) &  may be multiple   \\
\hline
HE~2149$-$2745           &                                            &           \\
Burud et al. (2002b)     & $\Delta(BA)$ = 103$\pm$12 days  (11$\%$)   & 65$\pm$8  \\
\hline\hline
{\bf COMBINED}           &                                            & {\bf 61$\pm$7} \\
\hline
\end{tabular}
\end{table}

Measuring  velocity   dispersions  in  faint  objects   in  not  easy,
especially when  they are  blended with bright  quasar images.  It has
been possible, so  far, only for a few  systems, but spectrographs are
starting  to  be  used  on  large  telescopes  where  adaptive  optics
observations  are carried  out  in a  flexible way.  Three-dimensional
spectrographs are planed  on these telescopes, and may  well yield not
only the velocity  dispersion of lenses, but also  their full velocity
field.  In many  cases,  one will  be  able to  measure  at least  the
velocity dispersion  profile, hence constraining much  better the mass
profile of the lens, independent of lensing.

\subsection{H$_0$ with \pg}

Although many lensed  quasars are potentially as powerful  as \pg\ for
constraining  H$_0$,  this nice  quadruple  remains  one  of the  best
studied cases so far, along with the double Q~0957+561.  Basically all
possible observational tools have been  used to study \pg, and we have
adopted  it as  an example  to  illustrate how  observations help  the
theorist in  quasar lensing.   Some of the  main results  obtained for
\pg\ are summarized  in Table 1, following the  chronological order of
the observational saga of the object.

As the observations of \pg\  were improving, various values were found
for H$_0$ along  the years.  This is mainly a  consequence of the mass
sheet degeneracy and  to the lack of observational  constraints on the
actual mass profile of the  lens.  Among the range of possible values,
the highest ones were  obtained for lenses with constant mass-to-light
ratios, and  lower values of H$_0$  were found for  models including a
dark matter halo.  If our knowledge  of the physics of galaxies is any
good, the  latter lens model is  closer to the truth  than the former.
This  results in  values for  H$_0$ that  disagree (see  Table~1) with
local  estimates, based on  standard candles,  i.e., H$_0$  = 72$\pm$8
\kmsmpc\ (Freedman et al. 2001).

\section{Mass Production of Time-Delays: H$_0$ with Other Lenses}

Measuring time-delays has long been  the main limitation to the use of
quasar  lenses in  cosmology.  Indeed,  obtaining on  a  regular basis
images of good  quality and for long period of time  is not easy. With
the  increasing number  of large  telescopes in  excellent  sites, and
operated in ``service mode'', we  have nevertheless entered a phase of
mass production of time-delays.  It  took more than a decade to obtain
the  time-delay  in Q~0957+561,  but  four  time-delays were  recently
measured  in  one  single  thesis  (Burud,  2000b)  thanks  to  modern
instrumentation  and  image  deconvolution techniques.   In  addition,
systematic imaging campaigns of lensed quasars such as the one carried
out by  CASTLE, facilitate  the modeling of  lens galaxies.   Deep HST
images are used in combination  with spectroscopy, and along with what
we know of the physics/dynamics of galaxies, to break the degeneracies
between the models.

Even if the  observational situation in quasar lensing  is expected to
improve a  lot in the coming  years, thanks to adaptive  optics and 3D
spectrographs,  we  can  already   say,  with  the  precision  of  the
observations collected  so far,  that lensed quasars  are at  least as
powerful as other methods to determine H$_0$.

Let us play the naive game that consists in taking from literature the
nine estimates  of H$_0$, for  each of the nine  measured time-delays,
and average them.  This is summarized in Table~2 for all known lenses,
along with the time-delay measurements.   We indicate in the table the
value  obtained for  H$_0$ with  all  kind of  different models,  just
scaling  the published  errors  to 1$\sigma$.   Some authors  consider
isothermal  spheres or  ellipsoids  with and  without external  shear.
Others prefer  to use lenses  with constant mass-to-light  ratios.  In
other  words, the  values  we quote  in  Table~2 are  all affected  by
unknown systematics.   These systematics  have been estimated  by each
author and  included in  the published error  bars.  If they  had been
under- or  over- estimated, the dispersion between  all the individual
estimates  of   H$_0$,  and  the   published  errors  on   one  single
measurement,  would be  incompatible.   Following our  naive idea  the
averaged value for H$_0$,  (without error weighting) is H$_0$=61$\pm$7
\kmsmpc. Figure~12 is  another way of displaying the  data of Table~2,
showing the  value of H$_0$ as  a function of the  lens redshift.  The
1$\sigma$ error region is dashed on the figure.  Four measurements (in
fact  3,  if  we include  the  mean  value  for  the 2  estimates  for
HE~1104-1805 in the dashed area)  are outside the 1$\sigma$ region, as
expected for  9 independent measurements.  Unless all  the authors did
the same  systematic error (and this  is unlikely because  of the very
different  choices of models),  the dispersion  between the  points is
fully compatible with the individual error bars.  This means that even
if the models suffer from  systematics, the amplitude of the error has
been correctly estimated by the authors.

\begin{figure}[t!]
\begin{center}
\leavevmode \includegraphics[height=12cm]{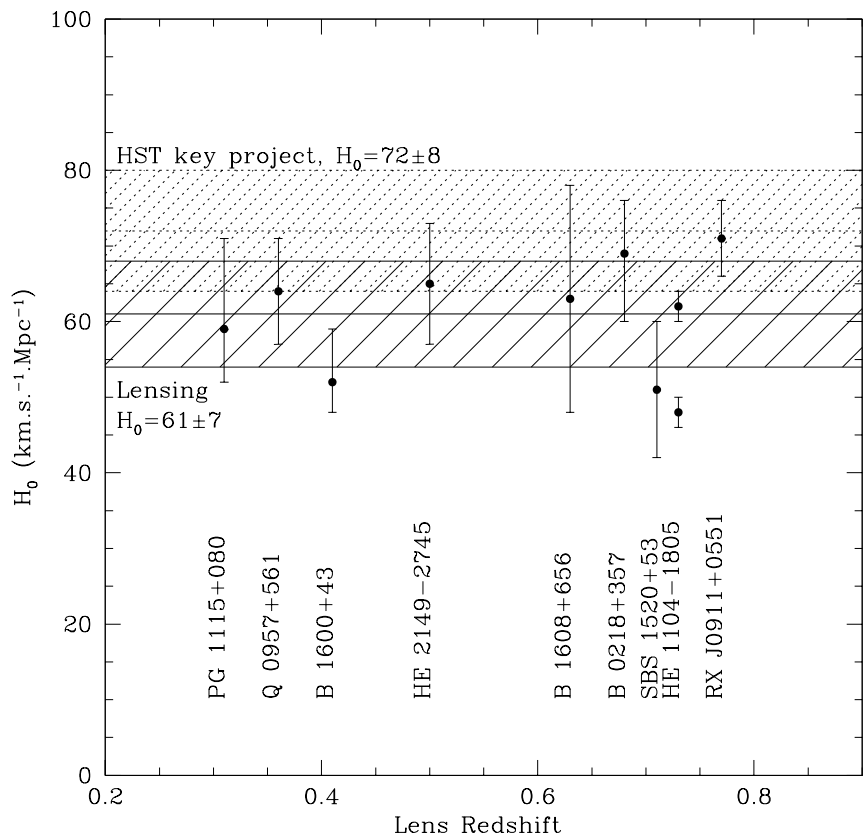}
\caption{Values of H$_0$ for  all lenses with known time-delays. H$_0$
is  given as  a function  of lens  redshift, directly  taken  from the
literature.   The dashed  area shows  the range  of  possible values,
according to lensing,  which is to be compared  with the dashed-dotted
area, infered from the Cepheid method  in the framework of the HST key
project.  As in the  case of  \pg, the  mean value  for H$_0$  is very
marginally  compatible   with  the   local  estimate  of   the  Hubble
parameter. }
\label{Ho}
\end{center}
\end{figure}

A more correct approach is to  impose that a given family of profiles
is used  to model the mass  distribution of lenses in  general, and to
impose that H$_0$ should be the same everywhere in the Universe.  This
has been  done so far by  very few authors.  Williams  \& Saha (2000),
use   two  systems   and   their  non-parametric   models,  to   infer
H$_0$=61$\pm$11 \kmsmpc  (1$\sigma$ error). Courbin et  al. (2002) add
two other  lenses to the Williams  \& Saha sample, and  adopt the same
non-parametric  technique  to  infer  H$_0$=64$\pm$4  \kmsmpc.   Using
parametric lens  models and  a set of  4 {\it different}  lenses (only
\pg\ is  common to all authors) Kochanek  (2002) obtain H$_0$=51$\pm$5
\kmsmpc  if lens galaxies  have dark  matter halos,  or H$_0$=73$\pm$8
\kmsmpc if lens galaxies have constant mass-to-light ratios.

Gravitationally lensed  quasars have  one drawback over  other methods
for  determining H$_0$:  they do  require good  knowledge of  the mass
profile  of  the  lens.   It   has  long  been  a  limitation  to  the
effectiveness of  the method to  produce reliable estimates  of H$_0$,
but this drawback is becoming easier and easier to overcome, thanks to
spectroscopy,  to high  spatial  resolution observations,  and to  the
progresses made on the physics of galaxies. Gravitational lenses have,
on the  other hand, tremendous  advantages over other  methods. First,
they  do not  rely on  any ``standard  candle'' which  would  rend the
method  close to  useless  if the  standard  candle turned  out to  be
significantly deviant from the ``standard'' behaviour. Second, it does
not require  secondary calibrators of  the standard candle,  e.g., low
redshift objects,  in the  case of the  supernovae method.   Third, it
probes  H$_0$  at cosmological  distances,  independent  of any  local
effect  such  as peculiar  velocities  of  nearby galaxies.   Finally,
lensing  is not  as  sensitive to  the  other cosmological  parameters
($\Omega_{m}$,  $\Omega_{\Lambda}$) than  the other  methods.

We compare in  Fig.~12, the value of H$_0$ for  lensed quasars and for
the Cepheid method. With the  present error bars, calculated with only
9   objects,   we   can   say   that  lensed   quasars   and   Cepheid
disagree. Lensing gives  lower values for H$_0$ than  does the Cepheid
method. If our  knowledge of the mass distribution  of galaxies is any
good, i.e.,  that galaxies have  dark matter halos, the  lensing value
would be even smaller.  Recent  CMB experiments such as WMAP find high
values for  H$_0$ (e.g.,  Spergel 2003), in  agreement with  the local
estimate: H$_0$=71,  with impressive error  bars of only  4$\%$.  With
the high degree of degeneracy  between the cosmological models used to
represent the observed  CMB power spectrum, H$_0$ is  in fact shown to
take any value between 54 and  72 (Efstathiou 2003).  It spans over an
even broader ranger if one does  not invoke any prior knowledge on the
value  of  $\Omega_{m}$  and  $\Omega_{\Lambda}$  as  determined  with
supernovae.

In fact,  no single method has been  proved to be good  enough that it
can surpass the  others.  There are, however, methods  that are better
than others as pinning down  {\it some} of the cosmological parameters
and it has  been emphasized many times (Bridle et  al.  2003) that all
methods should be combined in order to break the degeneracies inherent
to one specific method. Gravitational lensing is one of these methods.
It  has advantages and  drawbacks, but  probably many  more advantages
than drawbacks given the present observational context.

\acknowledgements The  author wishes to thank Alain  Smette and Pierre
Magain for  useful discussions.  Fr\'ed\'eric Courbin  is supported by
the European  Commission through a Marie  Curie Individual Fellowship,
under  grant  MCFI-2001-0242.   The collaborative  grant  ECOS/CONICYT
C00U05 between Chile and France is also gratefully acknowledged.

\label{page:last}
\end{document}